\documentclass[12pt,epsf]{article}
\usepackage{amssymb,amsmath}
\usepackage{graphicx}

\newcommand{\be}{\begin{equation}}
\newcommand{\ee}{\end{equation}}
\newcommand{\bea}{\begin{eqnarray}}
\newcommand{\eea}{\end{eqnarray}}
\newcommand{\bear}{\begin{eqnarray}}
\newcommand{\eear}{\end{eqnarray}}
\newcommand{\beas}{\begin{eqnarray*}}
\newcommand{\eeas}{\end{eqnarray*}}
\newcommand{\ba}{\begin{array}}
\newcommand{\ea}{\end{array}}
\newcommand{\adj}[1]{#1^\dagger}

\newcommand{\tr}{\operatorname{tr}}
\newcommand{\pd}[2][1]{\ifnum#1=1 \frac{\partial}{\partial {#2}} \else
  \frac{\partial^#1}{\partial {#2}^{#1}}\fi}
\newcommand{\dpd}[2][1]{\ifnum#1=1 \dfrac{\partial}{\partial {#2}} \else
  \frac{\partial^#1}{\partial {#2}^{#1}}\fi}
\newcommand{\td}[2][1]{\ifnum#1=1 \frac{d}{d{#2}} \else
  \frac{d^#1}{d{#2}^{#1}}\fi}

\def\identity{{\rlap{1} \hskip 1.6pt \hbox{1}}}
\newcommand{\nbox}{{\,\lower0.9pt\vbox{\hrule \hbox{\vrule height 0.2 cm \hskip 0.19 cm \vrule height 0.2 cm}\hrule}\,}}

\def\href#1#2{#2}

\begin{document}
\begin{titlepage}
\hfill
\vbox{
    \halign{#\hfil         \cr
           } 
      }  
\vspace*{20mm}
\begin{center}
{\Large \bf Momentum-space entanglement for interacting fermions at finite density}

\vspace*{16mm}
Ting-Chen Leo Hsu, Michael B. McDermott, Mark Van Raamsdonk
\vspace*{1cm}

{
Department of Physics and Astronomy,
University of British Columbia\\
6224 Agricultural Road,
Vancouver, B.C., V6T 1W9, Canada}

\vspace*{1cm}
\end{center}
\begin{abstract}
We investigate the entanglement between individual field theory modes in finite-density systems of interacting relativistic and non-relativistic fermions in one spatial dimension. We calculate the entanglement entropy for a single field theory mode and the mutual information between any two modes. The calculation is perturbative in the four-fermion (two-body) coupling, with the leading contribution at order $\lambda^2 \log(\lambda^2)$. At this leading order, the perturbative expression for the entanglement entropy of a mode diverges logarithmically as the momentum of the mode approaches the Fermi surface from above or below. The mutual information between modes is largest for pairs of modes just above and below the Fermi momentum. The entanglement properties of modes near the Fermi surface are qualitatively the same if the field theory is cut off to eliminate modes away from the Fermi surface.

\end{abstract}

\end{titlepage}

\vskip 1cm

\section{Introduction}\label{introsection}

The physics of finite density systems of interacting fermions plays a crucial role in our understanding of a wide variety of condensed matter systems, from ordinary materials to nuclear matter in a neutron star. To understand even crude macroscopic properties of these systems, quantum mechanics (e.g. the Pauli Exclusion Principle) is essential. Nevertheless, our theoretical investigations often focus on classical observables such as thermodynamic quantities, correlation functions, and response functions, since these quantities are simpler to access via experiment. In this paper, we instead introduce and investigate some intrinsically quantum observables in simple examples of finite-density fermion systems.

One of the key features that distinguishes quantum systems from classical systems is the possibility of entanglement between different degrees of freedom. This entanglement can be quantified: given any subset $A$ of degrees of freedom, the von Neumann entropy $S_A = -\tr( \rho_A \log \rho_A)$ of the density matrix $\rho_A = \tr_{\bar{A}}|\Psi \rangle \langle \Psi |$ provides a measure of the entanglement between $A$ and the rest of the system $\bar{A}$. This is known as the entanglement entropy of the subsystem $A$ (for a review, see \cite{nc}). Entanglement entropy and related observables have been studied extensively in quantum field theory and many-body systems over the past several years (see, for example \cite{Calabrese:2004eu,Casini:2004bw,Kitaev:2005dm,onemode}), but these studies typically choose $A$ to be the subset of degrees of freedom inside a particular spatial region. In this work, we study the entanglement entropy for even simpler subsystems: we take $A$ a subset of {\it momentum space}, focusing on the simplest possible subset consisting of a single field theory mode (i.e. a single allowed momentum).

Momentum-space entanglement entropy in quantum field theory systems was investigated in detail recently by Balasubramanian and two of the present authors in \cite{Balasubramanian:2011wt}. There, it was emphasized that entanglement entropy in momentum space vanishes in the ground state of non-interacting systems and remains finite in the continuum limit, in contrast to the position-space entanglement entropy which is non-zero for non-interacting systems and diverges in the continuum limit. The work \cite{Balasubramanian:2011wt} showed further that in the presence of weak interactions, the momentum-space entanglement entropy can often be computed in perturbation theory, as we review in section 2. The goal of this paper is to carry out such perturbative calculations of the entanglement entropy for the simplest possible interacting finite-density fermion systems, a gas of free non-relativistic or relativistic fermions in one dimension perturbed by a four-fermion (two-body) interaction.

We focus on two quantities in particular: the single-mode entanglement entropy $S(p)$, and the mutual information $I(p,q) = S(p) + S(q) - S(p,q)$  that measures entanglement and correlations between two individual modes at distinct momenta. In the non-relativistic case, for either lattice fermions (section 2) or continuum fermions (section 3), we show that the leading perturbative expression for $S(p)$ diverges logarithmically in $|p_F - p|$ as $p$ approaches the Fermi momentum from above or below. The logarithmic divergence also indicates a breakdown of perturbation theory when the momentum is within $e^{-c/ \lambda^2}$ of the Fermi momentum, where $c$ is some $\lambda$-independent constant. For continuum fermions, we also calculate the leading perturbative contribution to the mutual information between any two modes, both in the non-relativistic case and (in section 5) for Dirac fermions with a $(\bar{\psi} \psi)^2$ interaction. This quantity shows discontinuities when either of the momenta cross the Fermi surface and is largest when both momenta are near the Fermi point.

The renormalization group picture of such interacting fermion systems (see \cite{Shankar:1993pf} for a review) suggests that the low-energy physics of the systems we consider should be described by a scale-invariant Luttinger liquid system. Luttinger liquids correspond to stable RG fixed-points, so the low-energy physics should be largely insensitive to the details of physics for modes far from the Fermi surface. As a check of this, we show that the behavior of the entanglement entropy for modes near the Fermi-surface is the same in a theory with a cutoff $||k| - k_F| < \Lambda$. As discussed in \cite{Balasubramanian:2011wt}, there may be a direct connection between the behavior of systems under renormalization group flows and the momentum-dependence of entanglement observable; investigating this further is an interesting question for future work.

\section{Momentum-space entanglement in perturbation theory}

In this section, we review the general result \cite{Balasubramanian:2011wt} for entanglement entropy at leading order in perturbation theory and its application to the calculation of various measures of entanglement for subsystems corresponding to subsets of modes in momentum space.

\subsubsection*{General result for entanglement entropy in perturbation theory}

Consider any quantum system with Hilbert space ${\cal H} = {\cal H}_A \otimes {\cal H}_B$ and Hamiltonian $H = H_A \otimes \identity +  \identity \otimes H_B + \lambda H_{AB}$. Letting $|n \rangle$ and $|N \rangle$ be energy eigenstates of $H_A$ and $H_B$ respectively, a state $| n \rangle \otimes | N \rangle$ (an energy eigenstate of the $\lambda =0$ Hamiltonian, e.g. the vacuum state) has no entanglement between the subsystems. Turning on the interaction, the perturbed eigenstate may be calculated using ordinary quantum-mechanical perturbation theory. From this, we can compute the density matrix for subsystem $A$ and the associated entanglement entropy $S_A$. As shown in (\cite{Balasubramanian:2011wt}), the leading order perturbative expression in the non-degenerate case is
\be
\label{master}
S_A = - \lambda^2 \log(\lambda^2) \sum_{n' \ne n, N' \ne N} { |\langle n',N' | H_{AB} | n, N \rangle|^2  \over (E_n + \tilde{E}_N - E_{n'} - \tilde{E}_{N'})^2} + {\cal O} (\lambda^2) \, .
\ee
This involves a sum over matrix elements of the interaction Hamiltonian between the original state and states for which both subsystems $A$ and $B$ have been changed.

\subsubsection*{Entanglement entropy for a region of momentum space in quantum field theory}

For quantum field theory, we can start at finite volume so that the the Hilbert space has a discrete Fock-space decomposition ${\cal H} = \otimes_p {\cal H}_p$ and the unperturbed Hamiltonian is a sum of terms
\[
H_0 = \sum_p E_{p, \alpha} a^\dagger_{p, \alpha} a_{p, \alpha} \; ,
\]
each acting on a single factor of the tensor product. Here, $\alpha$ labels the species of particle if there is more than one, and $E_{p, \alpha}$ is the energy of a particle of species $\alpha$ with momentum $p$. This may include the contribution of a chemical potential added to give a ground state with finite density
\be
\label{state}
|\Psi_i \rangle = \prod_i a^\dagger_{p_i, \alpha_i} | 0 \rangle \; .
\ee

We can take $A$ to be some subset of the allowed momenta for the theory, i.e. all modes with a particular set of allowed wavelengths, and consider the entanglement entropy of the modes in region $A$ for the state (\ref{state}). Taking the large volume limit with the region $A$ of momentum space fixed, we find \cite{Balasubramanian:2011wt} that the formula (\ref{master}) gives an entanglement entropy for modes in the region $A$ that is extensive (i.e. proportional to spatial volume), with
\be
\label{master2}
S_A/V = - \lambda^2 \log(\lambda^2) \sum_f \int^* \prod  {d^d p_a \over (2 \pi)^d} (2 \pi)^d \delta(p_f - p_i) {|{\cal M}_{fi}|^2 \over (E_f - E_i)^2 }   + {\cal O} (\lambda^2) \, .
\ee
Here, the matrix element ${\cal M}_{fi}$ is defined by
\be
\label{matrix}
\langle \Psi_f | H_{I} | \Psi_i \rangle = (2 \pi)^d \delta(p_f - p_i) {\cal M}_{fi} (p_1, \dots, p_N) \;
\ee
where $|\Psi_f \rangle$ are occupation number basis elements of the form $a^\dagger_{p_1, \alpha_1} \cdots a^\dagger_{p_n, \alpha_n} | 0 \rangle$. The sum and integral are over the possible states $|\Psi_f \rangle$ appearing in the matrix element. Specifically, the sum is over the possible number of particles of each type present in the state $| \Psi_f \rangle$, while the integral is over the momenta $p_a$ of the particles that have been added/removed from the initial state $| \Psi_i \rangle$ to produce $| \Psi_f \rangle$, with the constraint that we have added/removed at least one particle with momentum in the region $A$ and at least one particle with momentum in the complementary region $\bar{A}$.

\subsubsection*{Single-mode entanglement entropy in quantum field theory}

In the case where $A$ corresponds to a single mode, the entanglement entropy is finite and volume-independent in the large volume limit (since we are no longer keeping the momentum-space volume of the the region $A$ fixed in the limit). We find
\be
\label{singlemode}
S(p) = - \lambda^2 \log(\lambda^2) \sum_f \int^* \prod  {d^d p_a \over (2 \pi)^d} (2 \pi)^d \delta(p_f - p_i) {|{\cal M}_{fi}|^2 \over (E_f - E_i)^2 }   + {\cal O} (\lambda^2) \, .
\ee
where now in the sum/integrals over the basis state $|\Psi_f \rangle$ the constraint is that the occupation number of the mode with momentum $p$ is different than for $|\Psi_i \rangle$, and the occupation number of at least one mode with some other momentum has changed.

For the special case of spinless fermions, the Hilbert space associated with a single mode is only two-dimensional, and this allows us to give an explicit result for the order $\lambda^2$ terms in the entanglement entropy in terms of the order $\lambda^2 \log(\lambda^2)$ piece. From equation (18) in \cite{Balasubramanian:2011wt}, we see that for
\[
S(p) = - \lambda^2 \log(\lambda^2) a + {\cal O}(\lambda^2),
\]
we must have
\[
S(p) = - \lambda^2 \log(\lambda^2) a + \lambda^2 a (1 - \log(a)) + {\cal O}(\lambda^3),
\]
Below, we will write explicitly only the leading ${\cal O}(\lambda^2 \log(\lambda^2))$ terms.

\subsubsection*{Mutual information between modes in quantum field theory}

Finally, we will consider the mutual information between two modes with momentum $p$ and $q$. Letting $A_p$ and $A_q$ represent subsets of momentum space corresponding to infinitesimal volumes $d^d p$ and $d^d q$ about momenta $p$ and $q$, we find that the mutual information $I(A_p, A_q) \equiv S(A_p) + S(A_q) - S(A_p \cup A_q)$ is proportional to volume in the large volume limit, and also proportional to $d^d p$ and $d^d q$. If we define ${\cal I} (p,q)$ by
\be
\label{mutual}
I(A_p, A_q)/V = {d^d p \over (2 \pi)^d} {d^d q  \over (2 \pi)^d} {\cal I} (p,q) \; ,
\ee
then we find
\be
\label{mutualeq}
{\cal I}(p,q) = - \lambda^2 \log(\lambda^2) \sum_f \int^* \prod  {d^d p_a \over (2 \pi)^d} (2 \pi)^d \delta(p_f - p_i) {|{\cal M}_{fi}|^2 \over (E_f - E_i)^2 }   + {\cal O} (\lambda^2) \, ,
\ee
where now the state $|\Psi_f \rangle$ in the matrix element is required to differ from the state $|\Psi_i \rangle$ in the occupation numbers of modes $p$ and $q$ and at least one other mode.

\section{Entanglement entropy for lattice fermions with nearest neighbor interactions}

We now study the entanglement between modes in several models of fermions at finite density, starting with a systems of spinless fermions on a lattice in one spatial dimension with nearest neighbor interactions.

We choose a Hamiltonian\footnote{Here, we have rescaled $H$ to be dimensionless. To restore the physical dimensions, we can multiply by $1/(a^2 m)$, where $a$ is the lattice spacing and $m$ is the effective particle mass, so that excitations with momentum $p \ll 1/a$ about the unfilled state have energy $p^2/(2m)$.}
\[
H = -{1 \over 2} \sum_j (\psi^\dagger_{j+1} \psi_j + \psi_j^\dagger \psi_{j+1}) + \lambda \sum_j (\psi^\dagger_j \psi_j - {1 \over 2})(\psi^\dagger_{j+1} \psi_{j+1}- {1 \over 2})
\]
for which the ground state has half-filling and an exact particle-hole symmetry \cite{Shankar:1993pf}. We define the momentum-space modes $\psi_p$ by
\[
\psi_p = \sqrt{a} \sum_{j} \psi_j e^{i j p a} \qquad \qquad \psi_j = \sqrt{a} \int_{-{\pi \over a}}^{\pi \over a} {dp \over 2 \pi} \; \psi_p e^{-i j p a}
\]
such that
\[
\{ \psi_p , \psi_q^\dagger \} = (2 \pi) \; \delta_{2 \pi \over a} (p - q) \; ,
\]
a delta function with period $2 \pi / a$, where $a$ is the lattice spacing.

The Hamiltonian becomes
\be
\label{Hzero}
H = \int_{-{\pi \over a}}^{\pi \over a} {dp \over 2 \pi} \; \psi^\dagger_p \psi_p (-\cos(pa))  + {\lambda \over 4} - \lambda \int_{-{\pi \over a}}^{\pi \over a} {dp \over 2 \pi} \; \psi^\dagger_p \psi_p + \lambda H_I
\ee
where
\be
\label{Hint}
H_I = - {a \over 8 \pi^3} \int_{-{\pi \over a}}^{\pi \over a} d P d Q d p d q \delta(P+Q-p-q) \psi^\dagger_{P} \psi^\dagger_{Q} \psi_{p} \psi_{q} e^{i (Q-q) a} \; .
\ee
The third term in $(\ref{Hzero})$ represents an adjustment to the chemical potential such that the state remains at half-filling in the presence of the interaction.

For $\lambda = 0$, the mode energy $-\cos(pa)$ is negative for $|p| <  \pi / (2a)$, so the ground state is
\[
|\Psi_i \rangle = \prod_{|p| <  {\pi \over 2a}} \psi^\dagger_p |0 \rangle \; .
\]
We can now calculate the entanglement entropy of a single mode with momentum $k/a$ using (\ref{singlemode}). In this case, all states $|\Psi_f \rangle$ for which the matrix element (\ref{matrix}) is nonzero have the same number of particles as $|\Psi_i \rangle$, with two particles removed inside the Fermi surface (at momenta $p$ and $q$), and two particles added outside the Fermi surface (at momenta $P$ and $Q$). Denoting such a state by $|\Psi_i ; P, Q, p, q \rangle$, we have
\[
\langle \Psi_i ; P, Q, p, q | H_I | \Psi_i \rangle = - a (2 \pi) \delta(P+Q - p -q)(e^{i(Q-q)a} - e^{i(Q-p)a} - e^{i(P-q)a} + e^{i(P-p)a})
\]
so that the squared matrix element appearing in (\ref{singlemode}) is
\[
|{\cal M}(P,Q,p,q)|^2 = a^2 |(e^{i Q a} - e^{i P a})(e^{i q a} - e^{i p a})|^2 \; .
\]
When the momentum $k/a$ is inside the Fermi surface, we can take $p=k/a$ and then integrate over all $q$ inside the Fermi surface and $(P,Q)$ outside the Fermi surface. Performing the $Q$ integral using the delta function sets $Q=k/a + q - P$ in the integrand and the remaining integral is over all possible $q$ inside the Fermi surface and $P$ outside the Fermi surface such that $Q=k/a + q - P$ is also outside the Fermi surface. The result is that the leading perturbative contribution to the entanglement entropy is
\[
S_{in}(k) =  - {1 \over 4 \pi^2}  \lambda^2 \log(\lambda^2) \int_{R_1} dq dP { 4 (\cos(P - k) - \cos(P-q))^2 \over (\cos(q) + \cos(k) - \cos(P) - \cos(k+q - P))^2},
\]
where the region of integration is
\[
R_1 = \{-k \le q \le {\pi \over 2} ; -\pi + {k + q \over 2} \le P \le -{\pi \over 2} \} \cup \{ -{\pi \over 2} \le q \le -k ; {\pi \over 2} \le P \le {k+q \over 2} + \pi \}\; ,
\]
and we have absorbed a factor of $a$ into the integration variables.

Similarly, when the momentum $k/a$ is outside the Fermi surface, we can take $P=k/a$, and integrate over $Q$ outside the Fermi surface and $(p,q)$ inside the Fermi surface. The result is
\[
S_{out}(k) =  - {1 \over 4 \pi^2} \lambda^2 \log(\lambda^2) \int_{R_2} dQ dp { 4 (\cos(p - k) - \cos(p-Q))^2 \over (\cos(Q) + \cos(k) - \cos(p) - \cos(k+Q - p))^2},
\]
where
\beas
R_2 &=& \{{\pi \over 2} \le Q \le \pi ; -{\pi \over 2} \le p \le {Q+k \over 2}-\pi \} \cup \{ -\pi \le Q \le -k ; -{\pi \over 2} \le p \le {k+Q \over 2} \}  \cr
&& \qquad \qquad \cup \{ -k \le Q \le -{\pi \over 2} ; {k+Q \over 2} \le p \le {\pi \over 2} \}\; .
\eeas
It is straightforward to show that $S_{out}(k) = S_{in}(\pi/2 - k)$; thus, the entanglement entropy is exactly symmetric about the Fermi surface, a consequence of particle-hole symmetry.

\begin{figure}
\centering
\includegraphics[width=0.5\textwidth,angle=270]{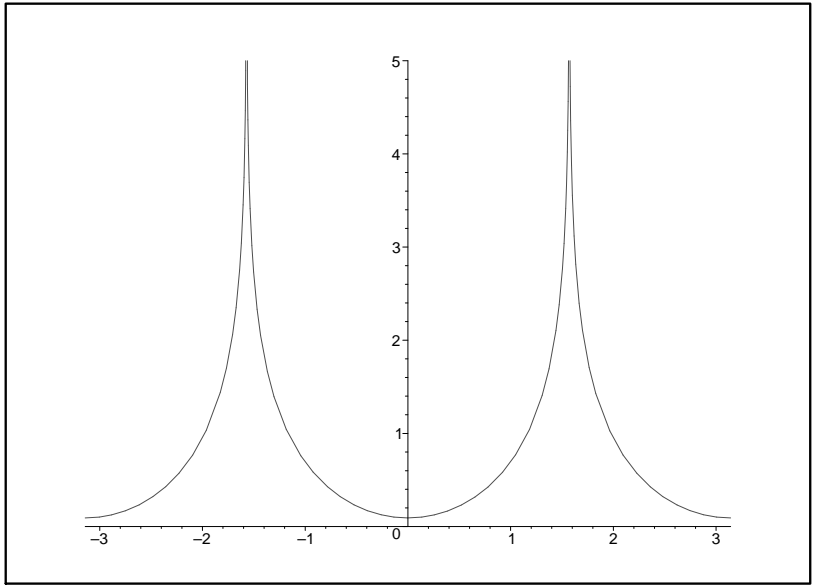}
\caption{Leading perturbative contribution to single mode entanglement entropy $S(k)$ as a function of mode momentum for lattice fermions at half filling. This diverges logarithmically at the Fermi points $k = \pm \pi/2$. }
\label{Slat}
\end{figure}

The function $S(k)$ is plotted in figure \ref{Slat}. We see that this leading perturbative expression diverges at the Fermi momentum; the divergence is logarithmic in $|k - k_F|$ and remains if we include a cutoff $|k - k_F| <  \epsilon$ restricting to momenta near the Fermi surface. In this case, we find
\[
S(k) = - {1 \over 4 \pi^2} \lambda^2 \log(\lambda^2) \ln \left({\epsilon \over |k - k_F|} \right) + C(\epsilon) + {\cal O}(k - k_F),
\]
where $C$ is a momentum-independent constant of order $\epsilon^2$.

Since the Hilbert space for a single mode is two-dimensional, the exact mode entanglement entropy is bounded by $\log(2)$. Thus, the divergence in our leading perturbative expression indicates a breakdown in perturbation theory when the momentum is taken too close to the Fermi surface. Specifically, we expect that the perturbative result is reliable only if it is much less than one. This requires that $|k - k_F| \gg e^{-1/ \lambda^2}$.

\section{Entanglement entropy for continuum non-relativistic fermions}

We now consider the continuum limit of the previous model, obtained by taking the lattice spacing to zero and adjusting the chemical potential so that states up to some fixed momentum (independent of $a$) remain occupied. Restoring the overall factor of $1/(a^2 m)$ in the Hamiltonian, we can rewrite the interaction (\ref{Hint}) as
\be
\label{Hint}
H_I = - {1 \over 32 \pi^3 a m} \int_{-{\pi \over a}}^{\pi \over a} d P d Q d p d q \delta(P+Q-p-q) \psi^\dagger_{P} \psi^\dagger_{Q} \psi_{p} \psi_{q} (e^{i Q a} - e^{i P a})(e^{i q a} - e^{i p a}) \; .
\ee
In the limit $a \to 0$, this gives
\be
\label{Hint}
H_I =  {a \over 32 \pi^3 m} \int_{-{\pi \over a}}^{\pi \over a} d P d Q d p d q \delta(P+Q-p-q) \psi^\dagger_{P} \psi^\dagger_{Q} \psi_{p} \psi_{q} (Q-P)(q-p) \; .
\ee
Rescaling $\lambda \to  \lambda / (a m)$ so that the Hamiltonian is independent of $a$ in the limit, we finally obtain
\[
H = H_0 + \lambda H_I
\]
where
\be
\label{Hzero1}
H_0 = \int_{-\infty}^{\infty} {dp \over 2 \pi} \; \psi^\dagger_p \psi_p ({p^2 \over 2 m} - \mu)
\ee
and
\be
\label{Hint1}
H_I =  {1 \over 32 \pi^3 m^2} \int_{-\infty}^{\infty} d P d Q d p d q \delta(P+Q-p-q) \psi^\dagger_{P} \psi^\dagger_{Q} \psi_{p} \psi_{q} (Q-P)(q-p) \; .
\ee
We now follow the same steps as in the previous section to obtain results for the single mode entanglement entropy. For $0 < k < 1$, we find
\[
S(p_F k) =  - {p_F^2 \over 4 \pi^2 m^2} \lambda^2 \log(\lambda^2) \left\{ \int_{-k}^{1} dq \int_{- \infty}^{- 1} dP + \int_{-1}^{-k} dq \int_{1}^{\infty} dP \right\} { (k-q)^2 (2P-k-q)^2 \over (P-k)^2 (P-q)^2 } \; .
\]
For $1 < k < 3$, we find
\[
S(p_F k) =  - {p_F^2 \over 4 \pi^2 m^2} \lambda^2 \log(\lambda^2) \left\{ \int_{-k}^{-1} dQ \int_{k + Q \over 2}^{1} dp + \int_{-2-k}^{-k} dQ \int_{-1}^{k + Q \over 2} dp \right\} { (k-Q)^2 (2p-k-Q)^2 \over (p-k)^2 (p-Q)^2 } \; .
\]
For $k>3$, we have
\[
S(p_F k) =  - {p_F^2 \over 4 \pi^2 m^2} \lambda^2 \log(\lambda^2) \left\{ \int_{-k}^{2-k} dQ \int_{k + Q \over 2}^{1} dp + \int_{-2-k}^{-k} dQ \int_{-1}^{k + Q \over 2} dp \right\} { (k-Q)^2 (2p-k-Q)^2 \over (p-k)^2 (p-Q)^2 } \; .
\]

\begin{figure}
\centering
\includegraphics[width=0.5\textwidth,angle=270]{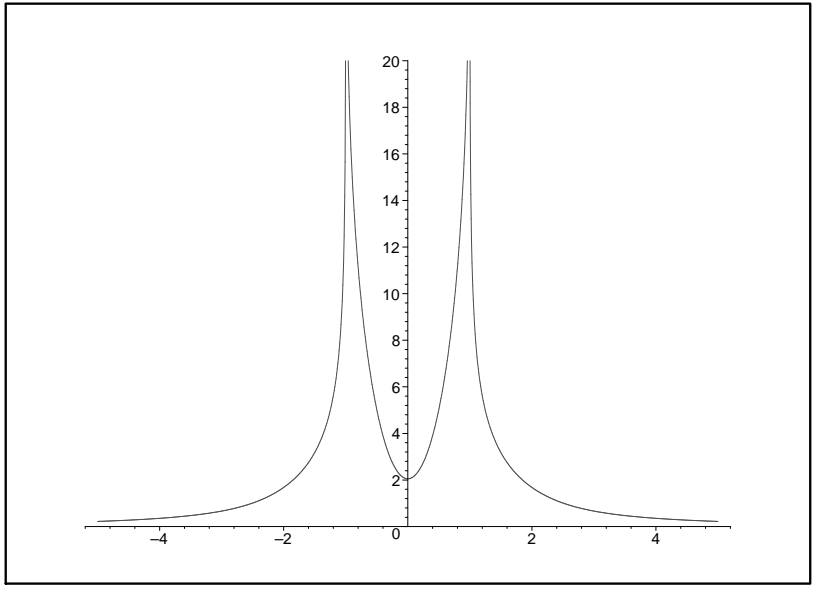}
\caption{Leading perturbative contribution to single mode entanglement entropy $S(p_F k)$ as a function of mode momentum (as a fraction of the Fermi momentum) for weakly interacting continuum non-relativistic fermions. This diverges logarithmically at the Fermi points. }
\label{scont}
\end{figure}

In each case, we have rescaled the integration variables by a factor of $p_F$ to make them dimensionless. All of these integrals may be evaluated analytically to obtain
\[
S(p_F k) =  - {p_F^2 \over 4 \pi^2 m^2} \lambda^2 \log(\lambda^2) f(k)
\]
where
\be
f(k) = \left\{ \ba{ll} - (k+1)^2 \ln(1-k)-(1-k)^2 \ln(1+k)  + k^2 \left({14 \over 3} + 2 \ln(2)\right) + {2 \over 3} + 2 \ln(2) & 0 < k < 1,\cr
-(k+1)^2 \ln(k-1) + (k+1)^2\left({7 \over 3} + \ln(2)\right) - {34 \over 3} (k+1) + {28 \over 3} - {8 \over 3 (k+1)} & 1 < k < 3 , \cr
{16 \over 3 (k^2 - 1)} & 3 < k .\ea \right.
\ee
The results for $k<0$ are obtained using $S(-k) = S(k)$.

The single-mode entanglement entropy $S(p_F k)$ is plotted in figure \ref{scont} . It diverges logarithmically at $k= \pm 1$ with the behavior on either side described by
\[
f(k) = -4 \ln|1-k| +{16 \over 3} {\rm sgn}(1-k) + 4 \ln(2) + {\cal O}(1-k).
\]
At $k=3$, the function $f(k)$ and its first, second, and third derivatives are all continuous, with a discontinuity appearing only in the fourth derivative.

\subsection{Mutual information between modes}

We can also look at the entanglement structure in more detail by calculating the mutual information between individual modes with momenta $p_F k$ and $p_F l$. Specifically, we calculate the function ${\cal I}(p_F k, p_F l)$ defined in (\ref{mutual}).

The mutual information ${\cal I}(p_F k, p_F l)$ satisfies
\[
{\cal I}(p_F k, p_F l) ={\cal I}(p_F l,p_F k)= {\cal I}(-p_F k, -p_F l) \; ,
\]
so we can restrict to the region $\{k>0, |l| < k\}$ and find ${\cal I}$ for the other values using the symmetries. For each choice of $k$ and $l$, the integral in (\ref{mutualeq}) is over distinct pairs of momenta such that together with the momenta $p_F k$ and $p_F l$, we have two momenta inside the Fermi surface and two momenta outside the Fermi surface (otherwise the matrix element in (\ref{mutualeq}) vanishes). Performing one integral using the delta function, we are left with a single integral in each case. For the various regions depicted in figure \ref{regions}, we find the following results:

\begin{figure}
\centering
\includegraphics[width=0.5\textwidth]{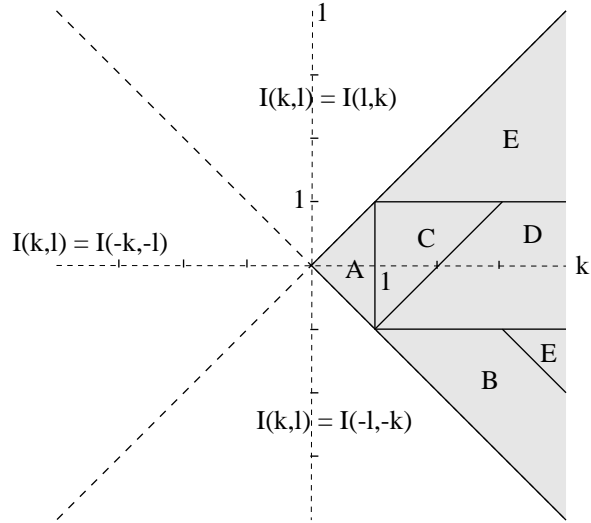}
\caption{Regions of $(k,l)$ space with different behaviors for ${\cal I}(k,l)$. Results for the unshaded regions may be obtained from the results for the shaded regions using the indicated symmetries of ${\cal I}$.}
\label{regions}
\end{figure}

\begin{figure}
\centering
\includegraphics[width=0.75\textwidth,angle=270]{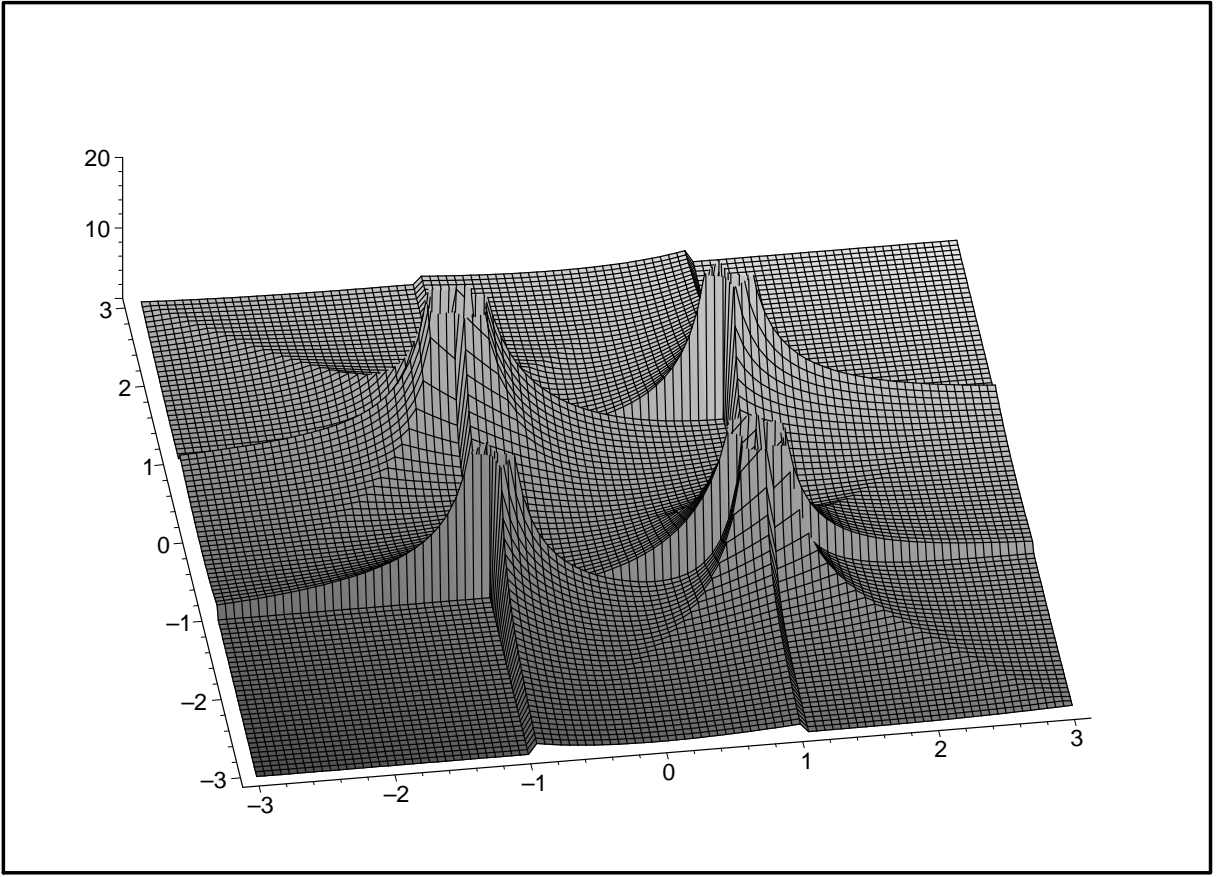}
\caption{Mutual information ${\cal I}(p_F k,p_F l)$ between individual modes for non-relativistic fermions with Fermi momentum $p_F$.}
\label{Ipq}
\end{figure}

\begin{figure}
\centering
\includegraphics[width=0.35\textwidth,angle=270]{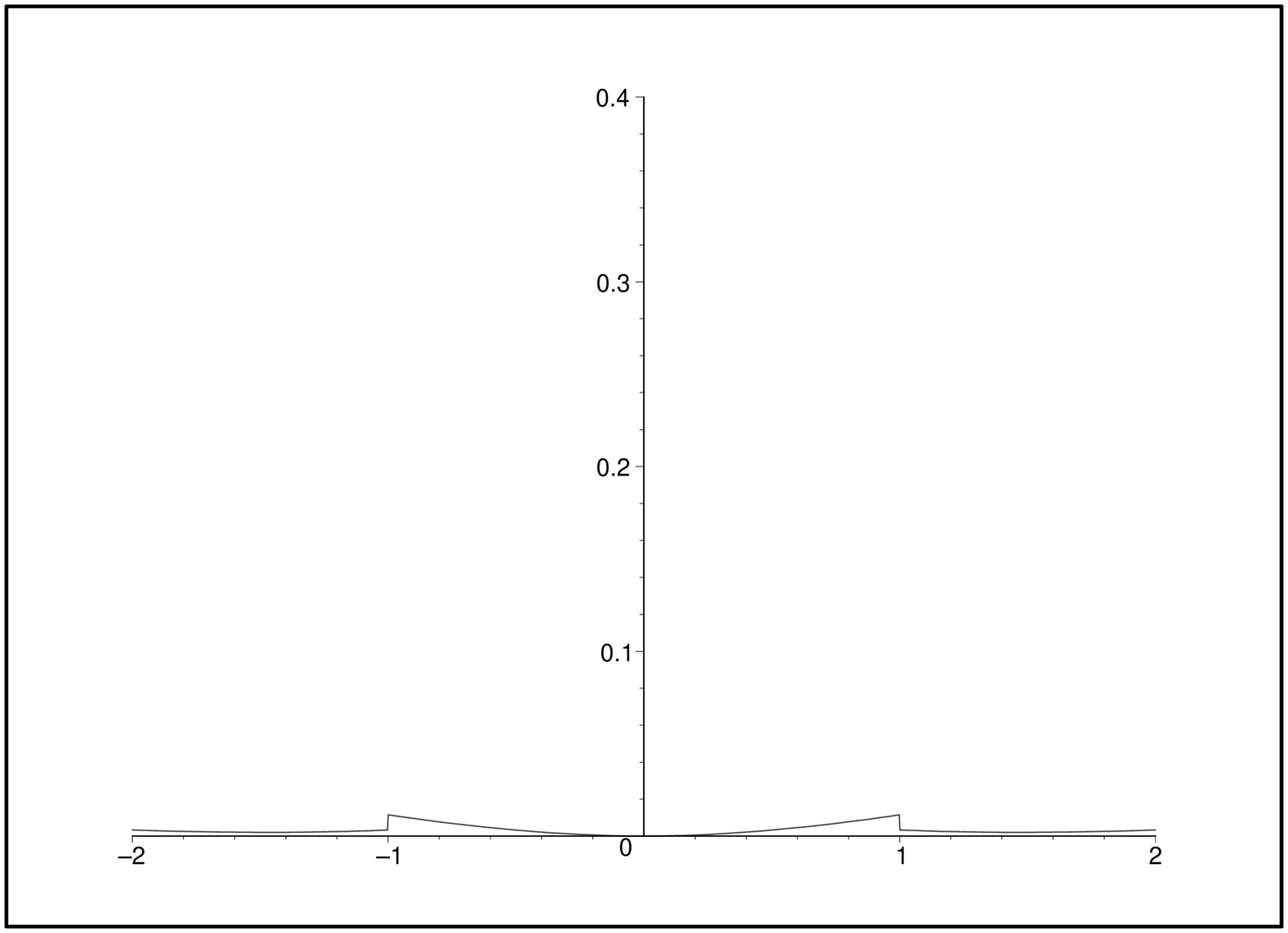} \includegraphics[width=0.35\textwidth,angle=270]{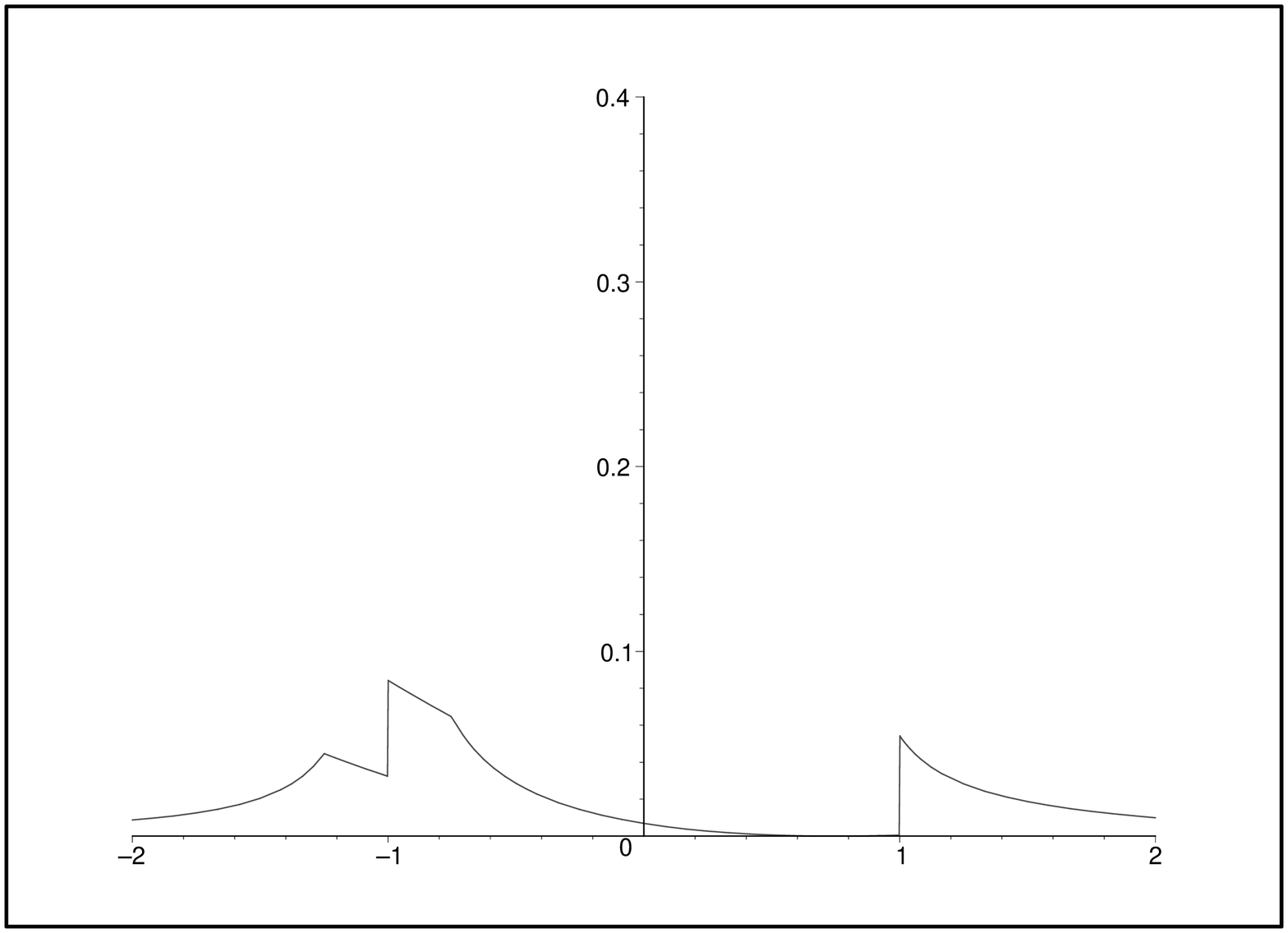} \\
\includegraphics[width=0.35\textwidth,angle=270]{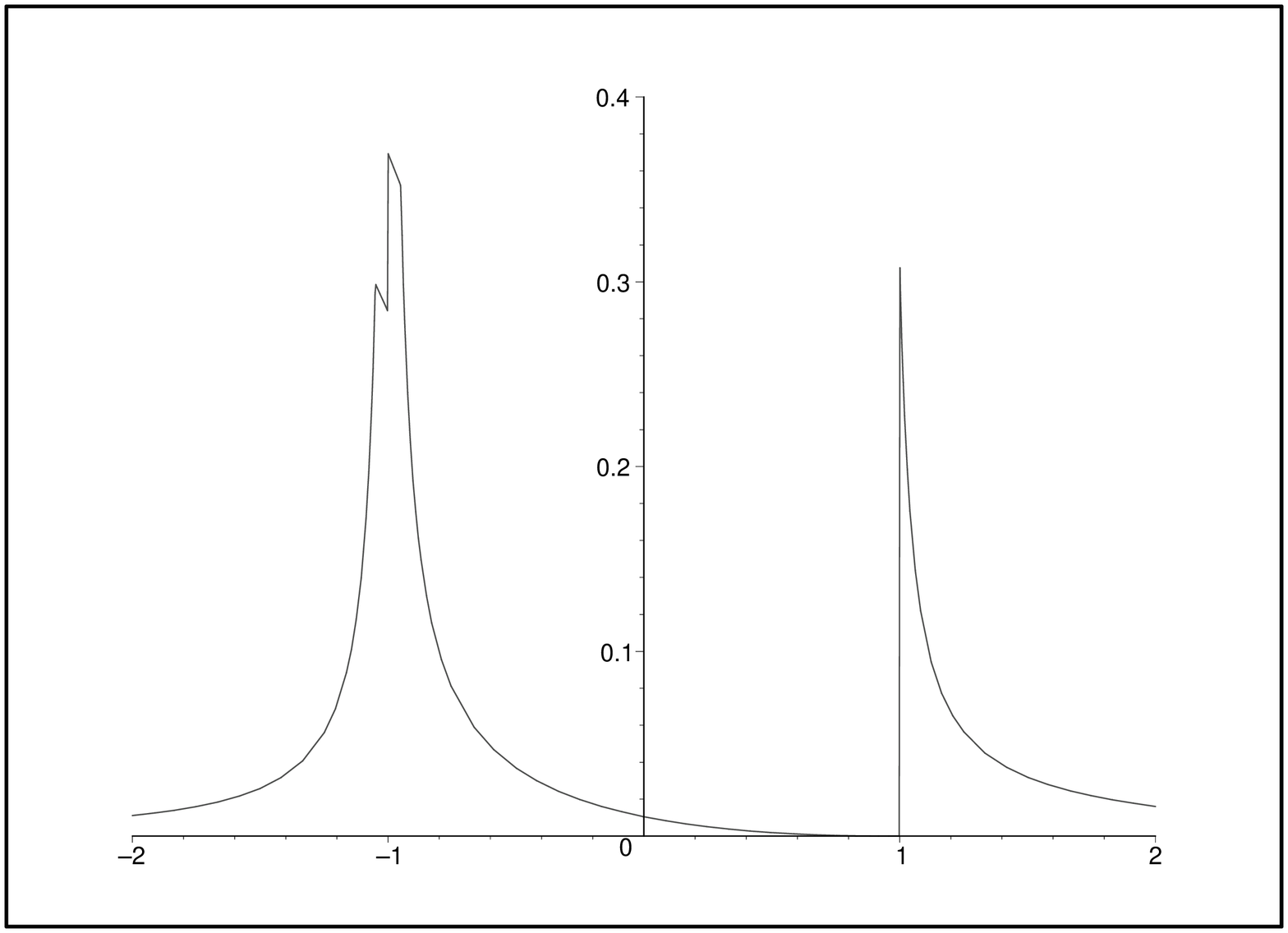} \includegraphics[width=0.35\textwidth,angle=270]{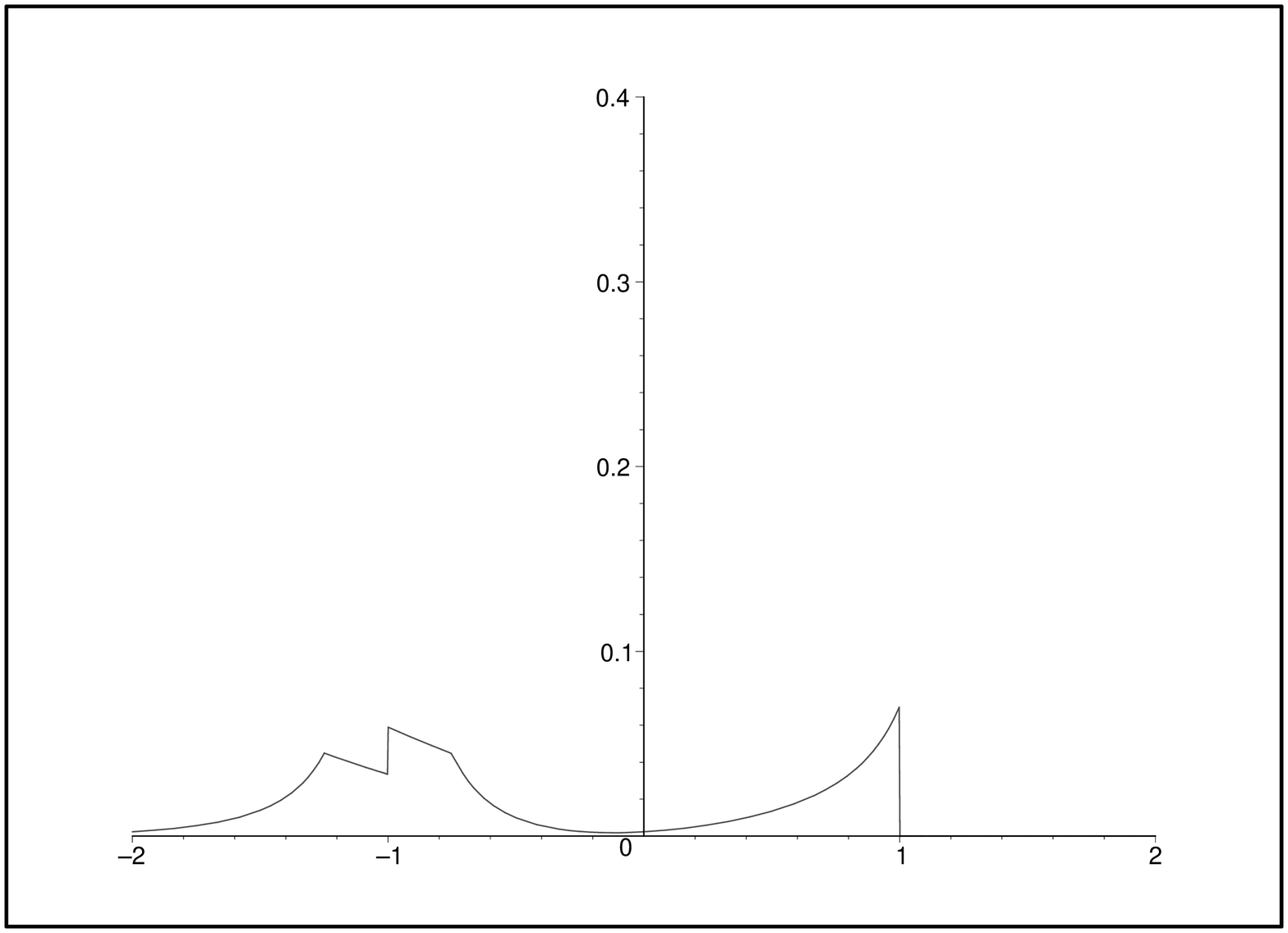}
\caption{Mutual information ${\cal I}(p_F k,p_F l)$ vs $l$ for non-relativistic fermions with $k=0$ (top left), $k=0.75$ (top right), $k=0.95$ (bottom left), and $k=1.25$ (bottom right). The overall scale for ${\cal I}$ is arbitrary}
\label{IpqNR}
\end{figure}

\begin{itemize}
\item
Region A: $\{0 < k < 1,  |l| < k \}$

\beas
{\cal I}(p_F k, p_F l) &=& - \lambda^2 \log(\lambda^2) {p_F \over m^2} \int_{-\infty}^{-1} {d P \over 2 \pi} {(k-l)^2 (2P - k - l)^2 \over (P-k)^2 (P-l)^2} \cr
&=& - \lambda^2 \log(\lambda^2) {p_F \over 2 \pi m^2} \left\{ 2(k-l) \ln \left( {1+k \over 1+l} \right) + {(k-l)^2 (k+l+2) \over (k+1) (l+1)} \right\}
\eeas
\item
Region B: $\{l < -1,  -l < k < 2-l \}$
\beas
{\cal I}(p_F k, p_F l) &=& - \lambda^2 \log(\lambda^2) {p_F \over m^2} \int_{k + l \over 2}^{1} {d p \over 2 \pi} {(k-l)^2 (2p - k - l)^2 \over (p-k)^2 (p-l)^2} \cr
&=& - \lambda^2 \log(\lambda^2) {p_F \over 2 \pi m^2} \left\{ 2(l-k) \ln \left( {1-l \over k-1} \right) + {(k-l)^2 (k+l-2) \over (k-1) (l-1)} \right\}
\eeas
\item
Region C: $\{-1 < l < 1,  1 < k < 2 + l \}$
\beas
{\cal I}(p_F k, p_F l) &=& - \lambda^2 \log(\lambda^2) {p_F \over m^2} \int_{-1}^{k-1-l} {d p \over 2 \pi} {(p-l)^2 (2k - p - l)^2 \over (k-p)^2 (k-l)^2} \cr
&=& - \lambda^2 \log(\lambda^2) {p_F \over 2 \pi m^2} \left\{ {(k-l)^3 \over (k+1)(l+1)} + {1 \over 3} {(k+1)^3 - (l+1)^3 \over (k-l)^2} - 2 (k-l) \right\}
\eeas
\item
Region D: $\{-1 < l < 1,  k > 2 + l \}$
\beas
{\cal I}(p_F k, p_F l) &=& - \lambda^2 \log(\lambda^2) {p_F \over m^2} \int_{-1}^{1} {d p \over 2 \pi} {(p-l)^2 (2k - p - l)^2 \over (k-p)^2 (k-l)^2} \cr
&=& - \lambda^2 \log(\lambda^2) {p_F \over 2 \pi m^2} \left\{ {1 \over 3} {(k+1)^3 - (k-1)^3 \over (k-l)^2} + {2(k-l)^2 \over k^2 -1} - 4\right\}
\eeas
Region E: $\{ k>1 , l > 1 \} \cup \{k > 3 , 2-k < l < -1 \}$
\[
{\cal I}(p_F k, p_F l) = 0 \; .
\]
\end{itemize}
The function ${\cal I}(p_F k, p_F l)$ is plotted in figure \ref{Ipq}, with plots for specific values of $k$ given in figure \ref{IpqNR}. We see that this leading-order contribution to ${\cal I}$ is generally discontinuous as one momentum crosses $\pm p_F$ and diverges when both momenta approach one of the Fermi points, unless the two momenta are equal.

\section{Entanglement entropy for relativistic fermions}

In this section, we consider Dirac fermions with a four-fermion interaction, described by the action
\[
S = \int d^2 x \left\{ i \bar{\psi} \gamma^\mu \partial_\mu \psi - m \bar{\psi} \psi - \tilde{\lambda} \bar{\psi} \psi \bar{\psi} \psi \right\} \; .
\]
This reduces exactly to the non-relativistic model in the previous section (with $\lambda = \tilde{\lambda}/2$) if we take chemical potential $\mu = m + \tilde{\mu}$ with $\tilde{\mu} \ll m$ and consider observables related to energy scales small compared to $m$ so that the particle number is fixed and antiparticles decouple.

Here, we consider general values of the chemical potential $\mu$ and the corresponding ground state $|\mu \rangle$ for which all particle states with energy less than $\mu$ are occupied (and none of the antiparticle states are occupied).

The field may be expanded as usual in terms of creation and annihilation operators as
\be
\label{exp}
\psi(x) = \int  {d k \over 2 \pi} \frac{1}{\sqrt{2\omega_k}}\left(a_k\, u_k\, e^{-ik\cdot x} + b^\dagger_k\, v_k\, e^{ik\cdot x}\right)\,.
\ee
where $a_k$ and $b_k$ respectively annihilate a particle and antiparticle with momentum $k$.

For the calculation of any of the entanglement observables in section 2, the non-vanishing matrix elements are those between the ground state and states obtained by
\begin{enumerate}
    \item adding two particles and two antiparticles, corresponding to the operator combinations $\adj{a}\adj{b}\adj{a}\adj{b}$ in the expansion of $\int dx\, (\bar{\psi} \psi)^2$;
    \item adding two particles and removing two particles, corresponding to $\adj{a}a\adj{a}a$; or
    \item adding two particles, removing one particle and adding an antiparticle, corresponding to $\adj{a}a\adj{a}\adj{b}$ and $\adj{a}\adj{b}\adj{a}a$.
\end{enumerate}
Particles can only be added outside the Fermi surface and can only be removed inside the surface; antiparticles can be added anywhere, but cannot be removed without annihilating the ground state.

Using (\ref{exp}), we can write the relevant terms in the interaction Hamiltonian in terms of the momentum-space creation and annihilation operators as
\beas
H_1 &=& \int \left\{ \prod {d k_i \over 2 \pi \sqrt{2 \omega_i}} \right\} \bar{u}(k_1) v(k_2) \bar{u}(k_3) v(k_4) a^\dagger_{k_1} b^\dagger_{k_2} a^\dagger_{k_3} b^\dagger_{k_4} (2 \pi) \delta(k_1 + k_2 + k_3 + k_4) \cr
H_2 &=& 2 \int \left\{ \prod {d k_i \over 2 \pi \sqrt{2 \omega_i}} \right\} \bar{u}(k_1) v(k_2) \bar{u}(k_3) u(k_4) a^\dagger_{k_1} b^\dagger_{k_2} a^\dagger_{k_3} a_{k_4} (2 \pi) \delta(k_1 + k_2 + k_3 - k_4) \cr
H_3 &=& \int \left\{ \prod {d k_i \over 2 \pi \sqrt{2 \omega_i}} \right\} \bar{u}(k_1) u(k_2) \bar{u}(k_3) u(k_4) a^\dagger_{k_1} a^\dagger_{k_3} a_{k_2} a_{k_4} (2 \pi) \delta(k_1 + k_3 - k_2 - k_4)
\eeas
Using these, we can calculate the matrix elements appearing in the calculation of entanglement observables. As an example, consider the matrix element of $H_1$ between the ground state $|\mu \rangle$ and a state $|\mu , p_1 p_2 ; p_3 p_4 \rangle$ where two particles with momenta $p_1$ and $p_2$ and two antiparticles with momenta $p_3$ and $p_4$ have been added to the ground state. We find
\be
\langle \mu , p_1 p_3 ; p_2 p_4 | H_1 | \mu \rangle = {(2 \pi) \delta (\sum p_i) \over 4 \sqrt{ \omega_1 \omega_2 \omega_3 \omega_4}} 2 (\bar{u}(p_1) v(p_2) \bar{u}(p_3) v(p_4) - \bar{u}(p_1) v(p_4) \bar{u}(p_3) v(p_2))
\ee
so that from (\ref{matrix}), we get
\beas
{|{\cal M}_{p_1 p_3 ; p_2 p_4}|^2 \over (\Delta E)^2} &=& {1 \over 4 \omega_1 \omega_2 \omega_3 \omega_4} {|\bar{u}(p_1) v(p_2) \bar{u}(p_3) v(p_4) - \bar{u}(p_1) v(p_4) \bar{u}(p_3) v(p_2)|^2 \over (\omega_1 +\omega_2 +\omega_3 +\omega_4)^2 }\cr
&=& {(p_1 \cdot p_3 - m^2)(p_2 \cdot p_4 - m^2) \over  \omega_1 \omega_2 \omega_3 \omega_4 (\omega_1 +\omega_2 +\omega_3 +\omega_4)^2} \cr
&\equiv& J_1(p_1,p_2,p_3,p_4)
\eeas

Similarly, we can calculate the matrix element of $H_2$ between the ground state and the state $|\mu , p_1 p_3 \bar{p}_2 ; p_4 \rangle$ where two particles and an antiparticle have been added with momenta $p_1$, $p_3$ and $p_4$ respectively, and a particle with momentum $p_2$ has been removed. We find
\beas
{|{\cal M}_{p_1 p_3 \bar{p}_2; p_4}|^2 \over (\Delta E)^2} &=& {1 \over 4 \omega_1 \omega_2 \omega_3 \omega_4} {|\bar{u}(p_1) v(p_4) \bar{u}(p_3) u(p_2) - \bar{u}(p_3) v(p_4) \bar{u}(p_1) u(p_2)|^2  \over (\omega_1 +\omega_3 -\omega_2 +\omega_4)^2}\cr
&=& {(p_1 \cdot p_3 - m^2)(p_2 \cdot p_4 + m^2) \over  \omega_1 \omega_2 \omega_3 \omega_4 (\omega_1 +\omega_3 -\omega_2 +\omega_4)^2} \cr
&\equiv& J_2(p_1,p_2,p_3,p_4) \; .
\eeas
Finally, we have
\beas
{|{\cal M}_{p_1 p_3 \bar{p}_2 \bar{p}_4}|^2 \over (\Delta E)^2} &=& {1 \over 4 \omega_1 \omega_2 \omega_3 \omega_4} {|\bar{u}(p_1) u(p_4) \bar{u}(p_3) u(p_2) - \bar{u}(p_3) u(p_4) \bar{u}(p_1) u(p_2)|^2  \over (\omega_1 +\omega_3 -\omega_2 -\omega_4)^2}\cr
&=& {(p_1 \cdot p_3 - m^2)(p_2 \cdot p_4 - m^2) \over  \omega_1 \omega_2 \omega_3 \omega_4 (\omega_1 +\omega_3 -\omega_2 -\omega_4)^2} \cr
&\equiv& J_3(p_1,p_2,p_3,p_4)
\eeas
for the matrix element of $H_3$ between the ground state and the state $|\mu , p_1 p_3 \bar{p}_2 \bar{p}_4 \rangle$ where particles have been added with momenta $p_1$ and $p_3$ and particles with momenta $p_2$ and $p_4$ have been removed.

\subsubsection*{Single-mode entanglement}

To calculate the entanglement entropy for a single mode with momentum $p$, we use the expression (\ref{singlemode}), taking the sum over the three types of final states discussed above. However, we find that the integrals in the terms involving $H_1$ and $H_2$ diverge. Thus, as noted in \cite{Balasubramanian:2011wt} for the case without chemical potential, the leading perturbative expression for the single-mode entanglement entropy in this model is ill-defined. Since the exact answer is necessarily less than $\log(2)$, this must indicate a breakdown of perturbation theory, as discussed in more detail in section 4.4 of \cite{Balasubramanian:2011wt}. Thus, for this model, we focus on the mutual information between modes, which can be computed in perturbation theory.

\subsubsection*{Mutual information between modes}

To calculate the mutual information between modes, we use (\ref{mutualeq}). In each case, ${\cal I}(p,q)$ is calculated using matrix elements for which the occupation number of the particle modes with momenta $p$ and $q$ have been changed relative to the ground state and for which occupation numbers for two other particles or antiparticles (with momenta $P$ and $Q$) have been changed.

When $p$ and $q$ are both inside the Fermi surface, we have:
\beas
{\cal I}(p,q) &=& -\lambda^2\log(\lambda^2)\frac{1}{2\pi}
\left[ {1 \over 2} \int_> d P \int_>  dQ \,  \delta(P + Q - p - q) J_3(P,p,Q,q)\right],
\eeas
where $\int_>$ and $\int_<$ indicates that the integration variable ranges outside and inside the Fermi surface, respectively.

When $p$ is inside, and $q$ is outside:
\beas
{\cal I}(p,q) &=& -\lambda^2\log(\lambda^2)\frac{1}{2\pi} \left[ \int_> d P \int dQ \,  \delta(P + Q - p + q) J_2(q,p,P,Q)  \right.\cr
&& \qquad \qquad \qquad \qquad \left. + \int_> d P \int_< dQ \,  \delta(P - Q - p + q) J_3(q,p,P,Q) \right].
\eeas
This also covers the case when $p$ is outside and $q$ is inside, since $I(p,q)=I(q,p)$.

Finally, when both $p$ and $q$ are outside:
\beas
{\cal I}(p,q) &=& -\lambda^2\log(\lambda^2)\frac{1}{2\pi}
 \left[ {1 \over 2} \int d P \int  dQ \,  \delta(P + Q + p + q) J_1(p,P,q,Q) \right. \cr
&& \qquad \qquad \qquad \qquad   + \int_< d P \int  dQ \,  \delta(-P + Q + p + q) J_2(p,P,q,Q)  \cr
&&  \left.\qquad \qquad \qquad \qquad \qquad  + {1 \over 2} \int_< d P \int_< dQ \,  \delta(-P - Q + p + q) J_3(p,P,q,Q) \right].
\eeas

\begin{figure}
\centering
\includegraphics[width=0.35\textwidth,angle=270]{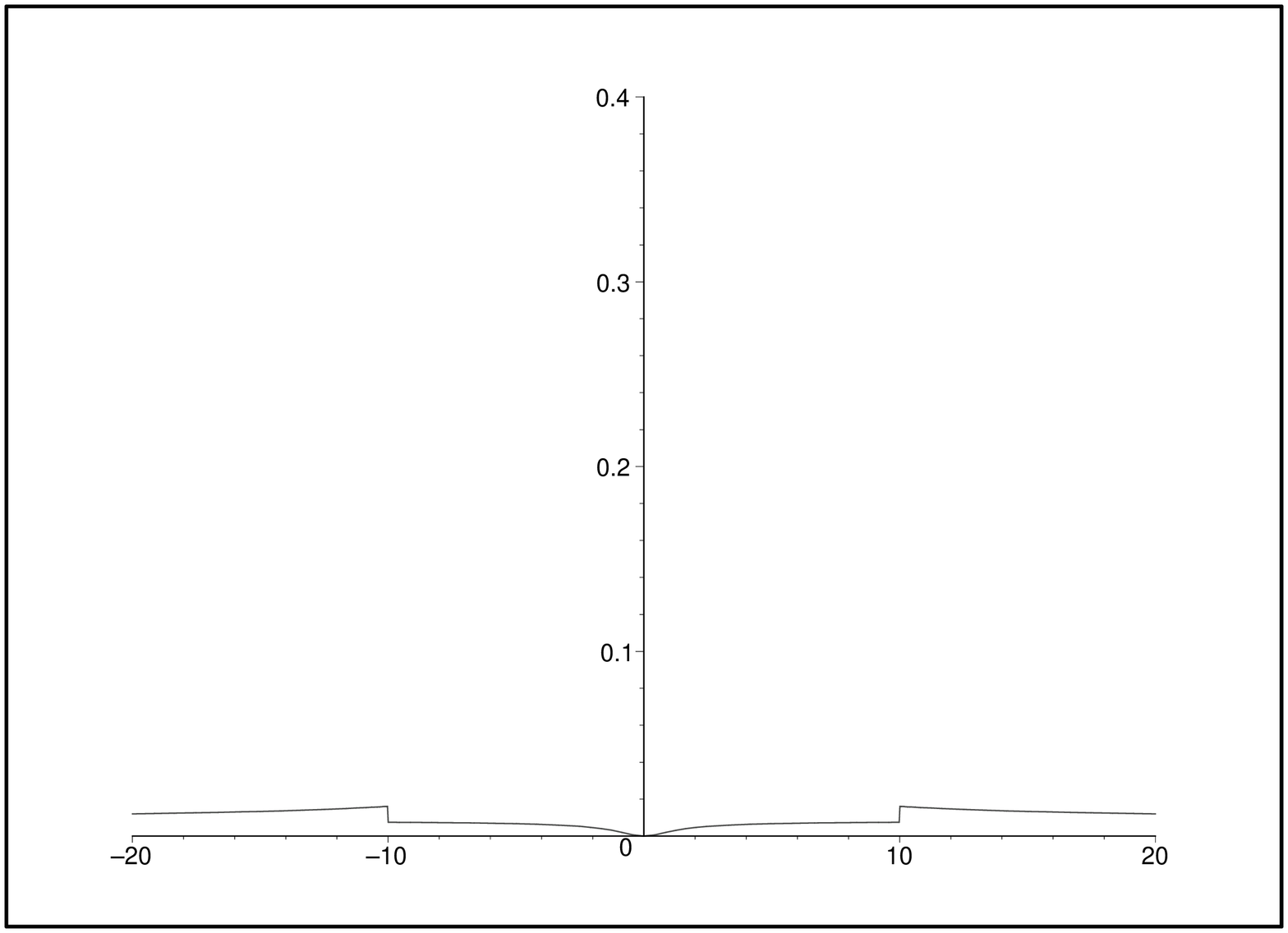} \includegraphics[width=0.35\textwidth,angle=270]{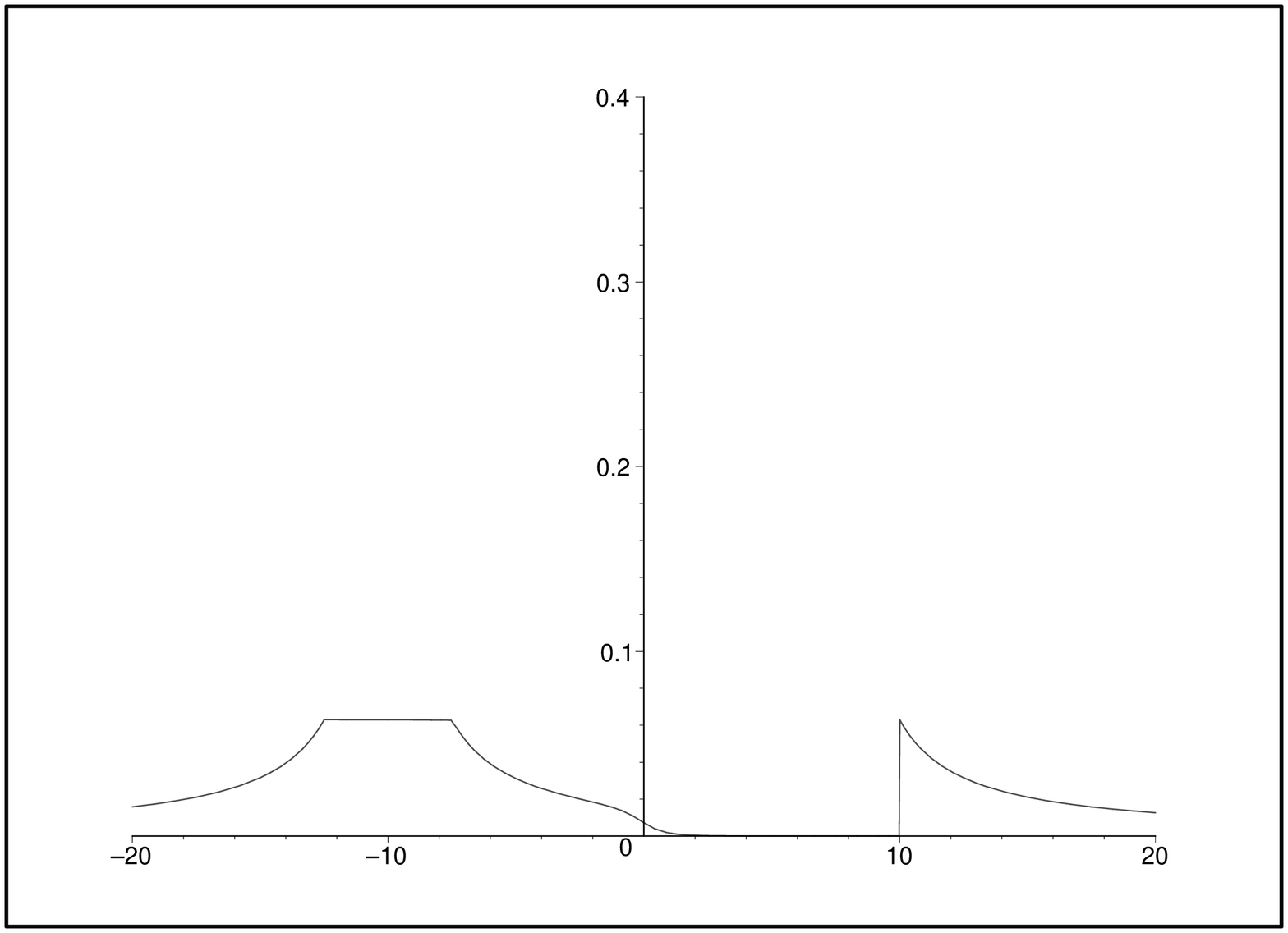} \\
\includegraphics[width=0.35\textwidth,angle=270]{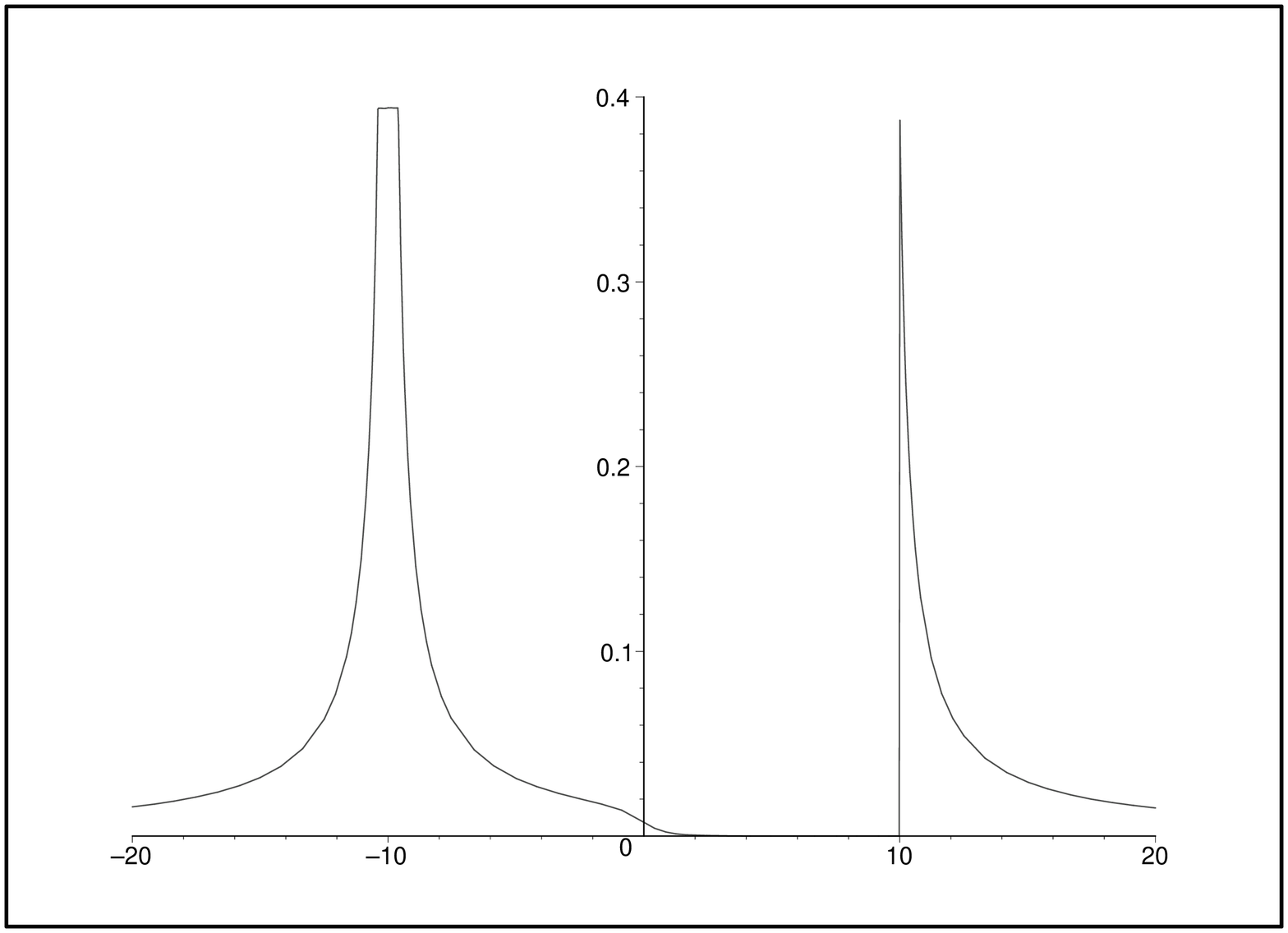} \includegraphics[width=0.35\textwidth,angle=270]{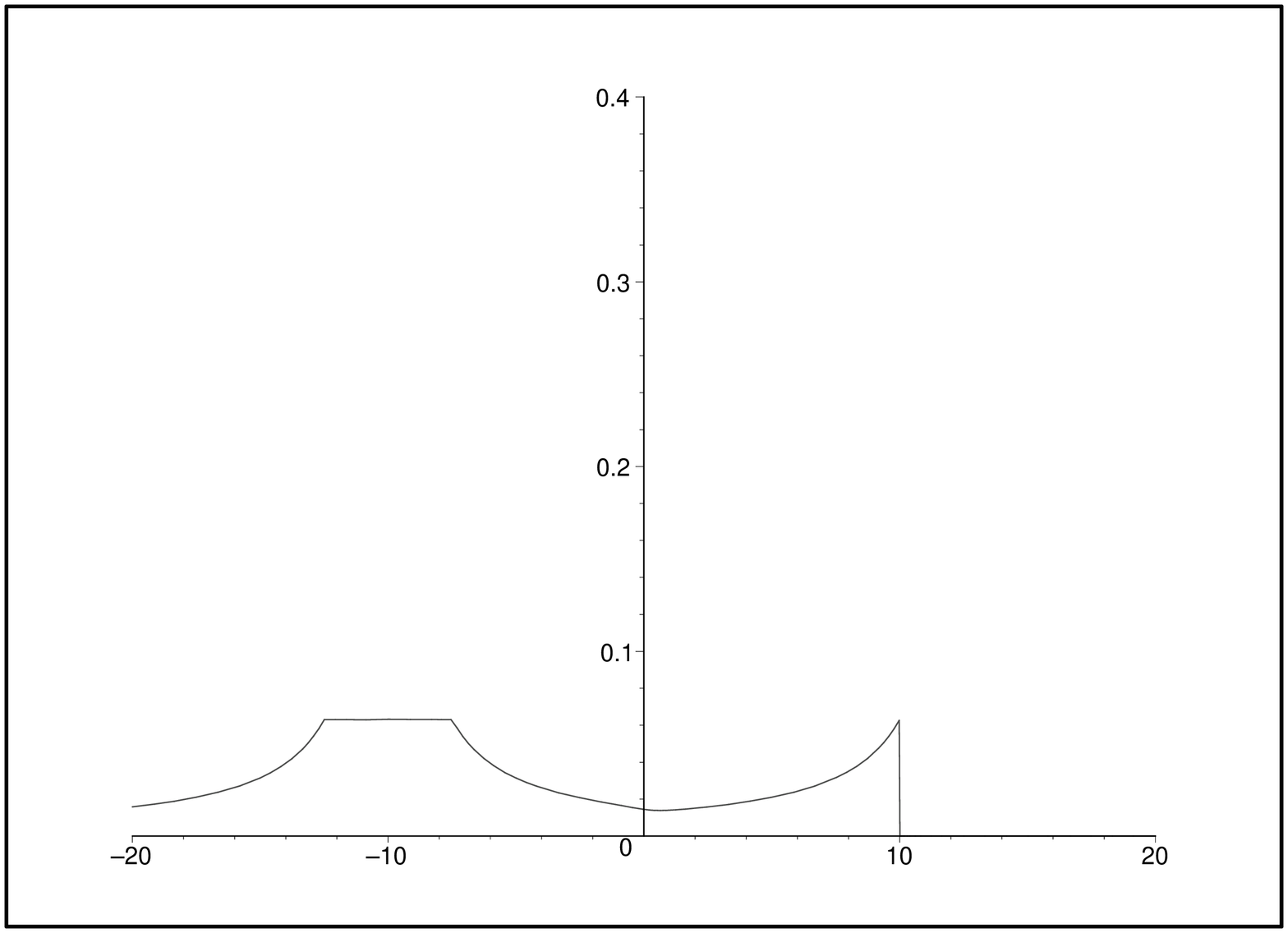}
\caption{Mutual information ${\cal I}(p,q)$ vs $q$ for relativistic fermions with $\mu=10m$ and $p=0$ (top left), $p=7.5$ (top right), $p=9.5$ (bottom left), and $p=12.5$ (bottom right). Momenta are given in units of $m$, so the Fermi points are at $\pm 10$. The overall scale for ${\cal I}$ is arbitrary.}
\label{Ipq_dirac10}
\end{figure}

All these integrals are straightforward to perform numerically. We find behaviour qualitatively similar to the non-relativistic case, with discontinuities at the Fermi momenta and the largest mutual information when both modes are close to the Fermi point. As an example, in figure \ref{Ipq_dirac10} the mutual information ${\cal I}(p,q)$ at $\mu = 10m$ is plotted as a function of $q$ for several fixed values of $p$. These may be compared with the non-relativistic results in figure \ref{IpqNR}.

\begin{figure}
\centering
\includegraphics[width=0.75\textwidth,angle=270]{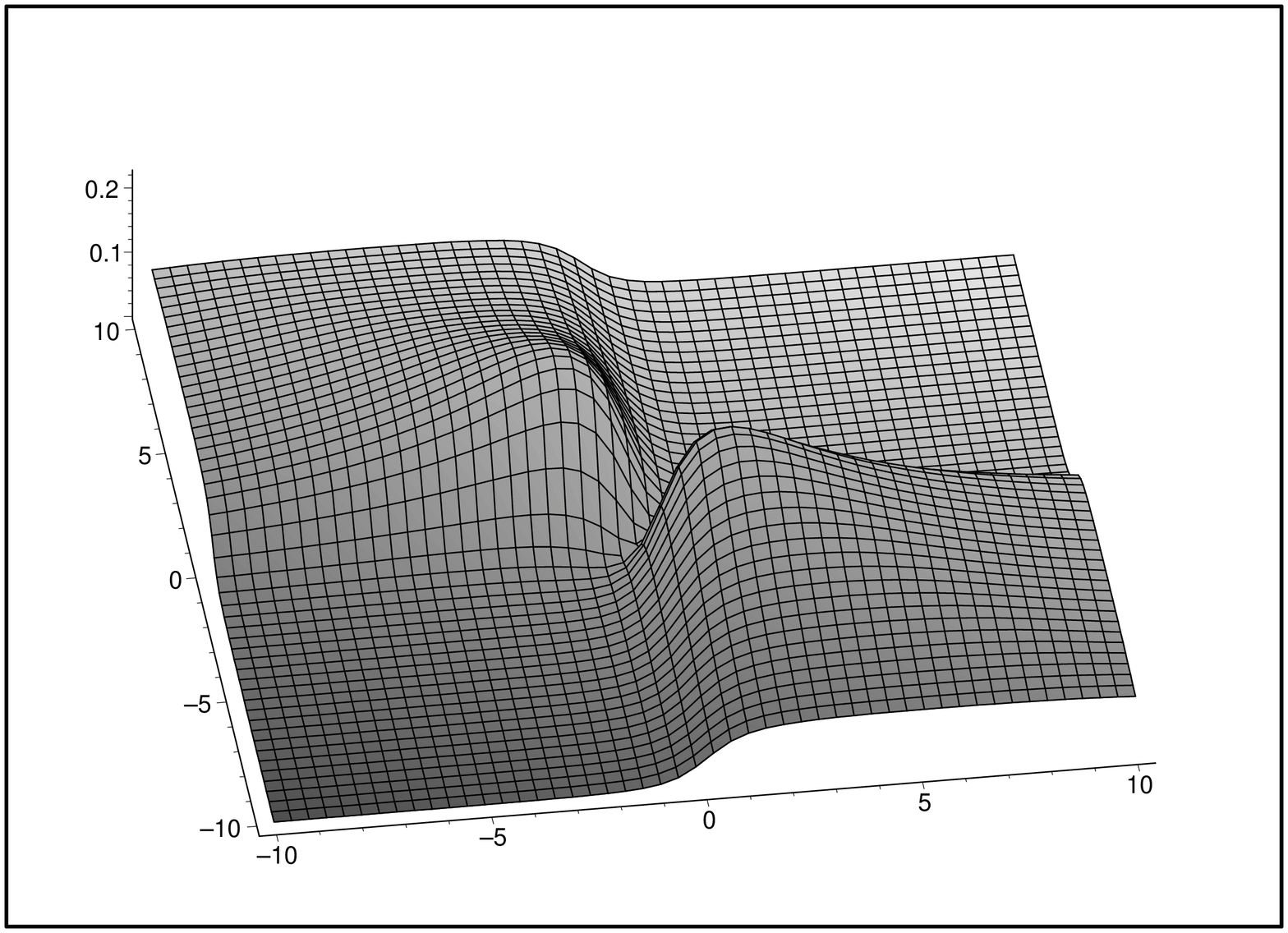}
\caption{Mutual information ${\cal I}(p,q)$ as a function of $p$ and $q$ for relativistic fermions with $\mu=0$. Momenta are given in units of $m$.}
\label{Ipq_dirac0}
\end{figure}

For comparison, we plot in figure \ref{Ipq_dirac0} the mutual information ${\cal I}(p,q)$ vs $q$ for several values of $p$ in the case with zero chemical potential, where the unperturbed ground state is the Fock-space vacuum. Here, we find smooth behavior with the largest mutual information between pairs of momenta of opposite sign and momenta of order the mass scale. The mutual information falls off as $1/q$ for fixed $p$.

\section*{Acknowledgments}

We are grateful to Ian Affleck, Vijay Balasubramanian and Albion Lawrence for helpful conversations. This research is supported in part by the Natural Sciences and Engineering Research Council of Canada and the Canada Research Chairs programme.

\end{document}